\documentclass[a4paper,12pt]{article}
\usepackage[left=0.9in,right=0.9in,top=0.9in,bottom=0.8in]{geometry}
\usepackage{amssymb}
\usepackage{amsmath}
\usepackage{physics}
\usepackage{graphicx}
\usepackage{float}
\usepackage{slashed}
\usepackage{appendix}
\usepackage[hidelinks]{hyperref}
\usepackage[sorting=none, style = phys, biblabel=brackets]{biblatex}
\usepackage{authblk}
\newcommand\blfootnote[1]{%
  \begingroup
  \renewcommand\thefootnote{}\footnote{#1}%
  \addtocounter{footnote}{-1}%
  \endgroup
}
\DeclareUnicodeCharacter{0301}{\'{i}}
\bibliography{KCpaper}
\newcommand{\be}{\begin{equation}}
\newcommand{\ee}{\end{equation}}
\newcommand{\bea}{\begin{eqnarray}}
\newcommand{\eea}{\end{eqnarray}}

\title{\textbf{Krylov construction and complexity for driven quantum systems}}
\author{Amin A. Nizami$^{*}$}
\author{Ankit W. Shrestha$^{\dagger}$}
 \affil{\small{Department of Physics, Ashoka University, Rajiv Gandhi Education City, Rai, NCR, India 131029}}
\date{}

\begin{document}

\maketitle
\begin{abstract}
Krylov complexity is an important dynamical quantity with relevance to the study of operator growth and quantum chaos, and has recently been much studied for various time-independent systems. We initiate the study of K-complexity in time-dependent (driven) quantum systems. For periodic time-dependent (Floquet) systems, we develop a natural method for doing the Krylov construction and then define (state and operator) K-complexity for such systems.  Focusing on kicked systems, in particular the quantum kicked rotor on a torus, we provide a detailed numerical study of the time dependence of Arnoldi coefficients as well as of the K-complexity with the system coupling constant interpolating between the weak and strong coupling regime. We also study the growth of the Krylov subspace dimension as a function of the system coupling constant.
    \blfootnote{ $^*$amin.nizami@ashoka.edu.in  \hspace{0.6cm} $^\dagger$sth.ankit61@gmail.com }
\end{abstract}
\section{Introduction}
The study of strongly coupled quantum many-body systems provides a variety of challenges and opportunities. These include physical questions related to understanding operator growth, thermalisation, scrambling of quantum information, as well as the structure and dynamics of quantum entanglement in many-body systems. Another important related topic is the study of quantum chaos. A reasonable and robust definition of quantum chaos (especially beyond the semi-classical regime) is still lacking, although various diagnostic measures have been proposed in the last few decades including the energy level spacing distribution, spectral form factors, Loschmidt Echo, out-of-time-order correlators (otoc) and circuit complexity \cite{BGS:84, Peres:1984pd, LO:69, Maldacena:2015waa}. There has also been a renewed interest in attempting to characterise chaotic dynamics via time-dependent quantities in quantum many-body systems which have no classical counterpart.

A relatively new quantity relevant to studies of operator growth and quantum chaos is Krylov complexity. Defined in \cite{Parker_2019} via a recursive algorithm (the Lanczos iteration, which we review in the next section) to study the growth of operator complexity, it depends on the Hamiltonian encoding the dynamics and the choice of the observable under study. The related state complexity was defined and studied in \cite{Balasubramanian_2022}.  A unified perspective on state and operator complexities was recently provided in \cite{Alishahiha:2022anw}.

The authors of \cite{Parker_2019} conjectured (and provided evidence for) the `Universal Operator growth Hypothesis' for general quantum systems in the thermodynamic limit.  This involves an upper bound on the rate of growth of the Lanczos coefficients  generated by the Lanczos algorithm and is related to the quantum chaos bound on the Lyapunov exponent in thermal systems \cite{Maldacena:2015waa}. The dynamics of the Lanczos coefficients and the related Krylov complexity has been numerically studied also for finite entropy sytems such as (integrable and non-integrable) spin-chains, Feingold-Peres and SYK models \cite{Parker_2019, integrability_to_chaos, Rabinovici_2021, Rabinovici_2022},  as well as the Bose-Hubbard model \cite{Bhattacharyya:2023dhp}. Krylov complexity has also been computed in field theoretic examples \cite{Dymarsky:2021bjq, Avdoshkin:2022xuw, Caputa:2021ori, Kundu:2023hbk}.

All these examples are of time-independent many-body Hamiltonian systems. However, time-dependent Hamiltonian systems provide some of the simplest settings to investigate (classical and quantum) chaos. As is well known, classical chaos manifests in a phase space of dimension at least three. Thus amongst the simplest chaotic systems are driven Hamiltonian systems with a single degree of freedom. These include one dimensional periodically kicked systems such as the quantum kicked rotor (QKR) \cite{Izrailev}. Such systems also have a tunable coupling constant, which can be used to interpolate between the weak and strong coupling domain. Due to the relative simplicity, it is thus desirable to study K-complexity in such systems before tackling the more complicated facets of {\it many-body} chaos.

The goal of this paper is to study K-complexity for time-dependent quantum systems. The main obstacle here is that the usual Krylov construction (reviewed below) - constructing the orthonormal Krylov basis using the Lanczos algorithm - works only for autonomous Hamiltonian systems. Focusing on periodically driven systems, also known as Floquet systems, we provide a natural and general Krylov construction based on a generalised Lanczos algorithm which can then be used to study (state and operator) Krylov complexity in such systems. Specialising further to kicked Floquet systems, we focus on a particular illustrative example, the QKR on a torus which is a finite dimensional, intrinsically quantum version of the canonical QKR (standard map on a cylinder) and investigate the dynamics of Arnoldi coefficients and K-complexity. Such quantum maps provide the simplest examples of physical systems manifesting quantum chaos and thus are ideal testing grounds for exploring the regular to chaotic transition as well as universal features of chaotic dynamics in the quantum domain.

Floquet systems, which will be the focus of our study, provide a fertile (experimental and theoretical) laboratory for investigating various facets of non-equilibrium quantum dynamics. The potential to engineer and control quantum states precisely and access new phases of matter away from thermal  equilibrium is one reason why the study of driven quantum systems is an important contemporary area of investigation. A whole host of rich physical phenomena arises in Floquet many-body systems. This includes the existence of novel out-of-equilibrium phases of matter including topological states, pre-thermalisation, many body localisation, dynamical quantum phase transitions and the existence of time crystals due to spontaneous breaking of time translation symmetry. For more details on these aspects, see the references \cite{Tsuji:2023dar, Zaletel:2023aej, d2013many, lazarides2014equilibrium, ponte2015many, abanin2017rigorous, harper2020topology, khemani2016phase, else2016floquet, oka2019floquet, kitagawa2010topological, yang2019floquet}.

This paper is structured as follows. In section \ref{overview} we give a brief overview of the basic notions of Krylov construction and K-complexity. Section \ref{KF}  explains how the construction works for general Floquet systems. Section \ref{TQKR} illustrates this method using the concrete examples of a toral QKR, and numerically explores the dynamics of the Arnoldi coefficients and K-complexity. Finally, section \ref{summary} provides a summary of the principal results and outlines future directions to be explored. The appendix outlines similar results obtained using the standard Lanczos algorithm with a Hermitian effective Floquet Hamiltonian.

\section{Overview of Krylov Construction and Complexity} \label{overview}
In this section, we provide a brief overview of the standard Krylov construction for time-independent Hamiltonians. For more details, see for example \cite{Parker_2019, geometry, Rabinovici_2022}.

\subsection{States}
We first review the notion of state complexity, also known as spread complexity \cite{Balasubramanian_2022}.  Consider a system with given (time-independent) Hamiltonian $H$ and initial state $\ket{\psi_0}$. We are interested in the time evolved state given by
\begin{align}
    \ket{\psi(t)}&= \exp (-itH) \ket{\psi_0}\nonumber \\ &=\ket{\psi_0}-itH \ket{\psi_0}+\frac{(-it)^ 2}{2!} H^ 2\ket{\psi_0}+....
\end{align}
so we construct the sequence of vectors $\{ \,\ket{\psi_0}, H \ket{\psi_0}, H^ 2 \ket{\psi_0}, H^ 3 \ket{\psi_0},....\,\}$ and then start the Gram-Schmidt like orthonormalisation process with $\ket{K_0}=\ket{\psi_0}$ and the recursive algorithm    
\begin{align}
    &\ket{K_n}=b_n^{-1} \ket{A_n},\nonumber \\ &\ket{A_{n+1}}=(H-a_n)\ket{K_n}-b_n \ket{K_{n-1}}
\end{align}
with $\ket{K_{-1}}=0$, and the Lanczos coefficients
\begin{align}
    a_n=\bra{K_n}{H}\ket{K_n}, \,\,\,\,\,\,b_n=\braket{A_n}^{1/2}.
\end{align}

Note that the Hamiltonian takes a tridiagonal (Hessenberg) form in the Krylov basis \footnote{The diagonal elements are the $a_n$'s, which are vanishing for the case of the operator Krylov construction discussed below.}. The state complexity is then defined as the expectation value of the K-complexity operator 
\begin{align} \label{KO}
\hat{K}=\sum_{n=0}^{K_D-1} n\ket{K_n}\bra{K_n},
\end{align}
in the time-evolved state
\begin{align}    
\mathcal{K}_{\psi}(t)= \bra{\psi(t)}\hat{K}\ket{\psi(t)}.
\end{align}

\subsection{Operators}
 The physical question of interest here is the growth of operator complexity and how to quantify it. To measure operator K-Complexity, we use a basis naturally generated by Heisenberg time evolution of operators.  
 An initially localized typical observable in a quantum many-body system generically spreads under time evolution:
\begin{align}   
O(t)=\exp (itH)O(0)\exp (-itH)= \exp(it \mathcal{L}_H) O(0) \nonumber\\
= O(0)+it [H,O(0)]+ \frac{(it)^ 2}{2}[H,[H,O(0)]]+.... \nonumber
\end{align}  

for any time-independent Hamiltonian $H$. The  following recursive construction is then natural: 
\begin{align}
\mathcal{L}_H|O_0)= |[H,O_0]),\,\, \mathcal{L}_H^ 2|O_0)= |[H,[H,O_0]]),\,\,....  \nonumber
\end{align}

and so on. Here we have defined an operator ket $|O_0)$, which is natural as the space of all operators acting on a vector space forms a vector space by itself. We have also defined the {\it Liouvillian super-operator} $\mathcal{L}_H$, which is an operator acting (as the adjoint action of $H$) on this space of operators. We also make the simple choice of the Frobenius inner-product on the space of operators\footnote{This induces the standard Hilbert-Schmidt operator norm. For thermal systems, a variant of this - the Wightman inner product - is more natural, see \cite{Parker_2019}.}: $(A|B)=\dfrac{1}{D}Tr(A^ {\dagger}B)$. We now choose the maximal linearly independent set from $\{|O_0), \mathcal{L}_H|O_0),\mathcal{L}_H^ 2 |O_0),.... \}$  and construct an orthonormal basis (Krylov basis) for this {\it Krylov subspace} using the following algorithm. Let the initial normalised state be $|K_0)=|O_0)$ and define $|K_1)=|[H,O_0])/b_1$ and continue recursively
\begin{align} \label{Lbasis}
|K_n)=\frac{1}{b_n}\big[\mathcal{L}_H |K_{n-1})-b_{n-1}|K_{n-2}) \big]
\end{align} 
Here, $b_n$ is the norm of the state in the r.h.s. numerator above. These normalisation coefficients encode the dynamics of the system and are called Lanczos coefficients. With the above choice of the inner product, one can explicitly check the orthonormality of the basis generated above.

If the system Hilbert space has dimension  $dim(\mathcal{H})=D$, the space of operators has $dim(\mathcal{H_O})=D^2$. 
The Krylov space is the minimal subspace  that contains the time evolution of $\mathcal{O}$ for all times \footnote{In other words it is the minimal invariant subspace of the Liouvillian that contains $\mathcal{O}(0)$.}. We will denote its dimension by $D_K$. It was shown in \cite{Rabinovici_2021} that $1\leq D_K\leq D^2-D+1$. Generic initial states/operators in chaotic systems are expected to lead to saturation of this upper bound whereas spectral degeneracies of the Liouvillian and system symmetries usually reduce the Krylov space dimension \cite{Rabinovici_2021, Rabinovici_2022}.
In this context, by generic  we mean that the initial state/operator has a non-zero projection on all the Liouvillian eigenvectors.

It is also possible to map this general quantum system in the Krylov basis representation to a particle hopping model on a discrete chain, where $b_n$ is the hopping amplitude at the $n$-th site. By writing $|O(t))=\sum_n i^n \phi_n(t)|K_n)$ and using the Heisenberg equation of motion for the operator together with the definition of the Lanczos algorithm (eq. \eqref{Lbasis}), we obtain a Schrodinger-like equation for the operator amplitudes $\phi_n(t)$:
\begin{align} \label{KCh}
\dot{\phi}_n(t)= b_n \phi_{n-1}(t)-b_{n+1}\phi_{n+1}(t)
\end{align}
with $\phi_{-1}(t)=0$ and $\phi_n(0)=\delta_{n0} $ . See, for example, \cite{Parker_2019, geometry, Rabinovici_2022} for more details.

K-complexity is then defined as 
\begin{align}   
\mathcal{K}_O(t)=\sum_{n=0}^{D_K-1} n (O(t)|K_n)(K_n|O(t))=\sum_{n=0}^{D_K-1} n |\phi_n(t)|^2
\end{align}
and is the average particle position on the Krylov chain.

It is clear from this definition that $D_K$ provides an upper bound on the Krylov complexity \footnote{In fact the stronger upper bound of $D_K/2$ was demonstrated in \cite{Rabinovici_2021}.}.
In the Krylov chain picture, the dimension of the linear subspace containing the time evolution of the operator is given by the effective size of the chain. Underpinning this idea of operator complexity is the characterisation of chaotic behaviour through the saturation value of K-complexity (average position on the chain at late times) \cite{Rabinovici_2022, Rabinovici_2021}.

Other than the average position on the chain, one can also compute the K-entropy \cite{Rabinovici:2019wsy} - the dispersion about the mean. This is the Shannon entropy of the distribution $\phi_n(t)$,
\begin{align}
S_K(t)=-\sum_{n=0}^{D_K -1} |\phi_n(t)|^2 \log |\phi_n(t)|^2 
\end{align}
and determines how randomized the distribution is (the extent of delocalisation on the Krylov chain).

It is noteworthy that K-complexity depends only on the Hamiltonian and the initial state/operator. These are precisely the two quantities that control classical chaos - parameters (in $H$) and phase space initial conditions \footnote{The one caveat to this is that for the case of operator K-complexity, it also further depends on the choice of the inner product on the vector space of operators and it is unclear if this is, in general, dictated solely by physical considerations.}.
  While this is merely suggestive and does not demonstrate the utility of K-complexity as a chaos diagnostic, one can contrast this with other quantities such as the Loschmidt echo, which is sensitive to how the Hamiltonian is perturbed, or the circuit complexity,
which has ambiguities such as the tolerance, choice of elementary gates and choice of the metric on the operator manifold.

The authors of \cite{Parker_2019} conjectured distinguishing behaviour of the Lanczos coefficients $b_n$'s for maximally chaotic systems in the thermodynamic limit
  \begin{align}
  b_n \sim \alpha n +\gamma+\mathcal{O}(1)
  \end{align}
leading to an exponential growth in K. Here the {\it growth rate} $\alpha >0$, as well as $\gamma$ are real constants. For the integrable case there is a softer growth $b_n \sim n^ {1/\delta}$ with $ \delta >1$. The chaos bound \cite{Maldacena:2015waa} on the quantum Lyapunov exponent takes the form $\lambda_L \le 2 \alpha$.  The reader may refer to \cite{Parker_2019} for further details.

In \cite{Rabinovici:2019wsy} the growth of operator complexity was investigated beyond the scrambling time. For finite dimensional chaotic systems with $f$ degrees of freedom,  the K-complexity shows an exponential growth till the scrambling time ($\mathcal{O}(\log f)$), and then transitions to a sustained period of linear growth until the Heisenberg time ($\mathcal{O}(e^{f})$) when it saturates to a value below $D_K/2$. One caveat to be noted here is that the initial exponential growth is a necessary but not sufficient condition for chaotic behaviour, because it can also be the signature of an unstable fixed point in phase space of an otherwise integrable system. This ``scrambling but not chaos" has been studied for K-complexity \cite{Bhattacharjee:2022vlt}  as well as the OTOC \cite{Xu:2019lhc, Dowling:2023hqc}

In \cite{Rabinovici_2021} it was argued that $D_K$ will be maximal (that is, $D_K=D^ 2-D+1$) for 1) a system without degeneracies in the spectrum of the Liouvillian and, 2)  the initial operator having a non-zero projection on all the Liouvillian eigenvectors. It was argued that generic operators in chaotic systems saturate this upper bound whereas for integrable systems $D_K$ is expected to be significantly smaller \footnote{For a 1D simple harmonic oscillator, with the position operator used to initiate the Krylov construction, the Krylov space is two dimensional even though $\mathcal{H}$ is infinite dimensional. However there are  exceptions to this, such as the strongly interacting integrable XXZ spin chain (solvable via the Bethe ansatz) where $D_K$ is maximal \cite{Rabinovici_2022}. We thank the referee for pointing this out.}. We will demonstrate that this is indeed the case for a finite dimensional quantum system (toral QKR) with a tunable coupling constant, which allows access to weak and strong coupling regimes.

\section{Krylov construction for Floquet systems} \label{KF}

In this section, we provide a quite natural and completely general method for the Krylov construction in any periodically driven time-dependent quantum system. This utilises the Floquet operator and generalises the usual Lanczos algorithm for constructing the Krylov basis. The method uses what is known as the Arnoldi iteration for constructing the Krylov basis. The operator Krylov construction for Floquet systems using this method was developed \footnote{We thank A. Dymarsky for bringing  this work to our attention after our arXiv preprint appeared.} in \cite{Yates:2021asz}. The Arnoldi iterative algorithm has recently \cite{Bhattacharya_2022, Bhattacharjee:2022lzy} also been utilised in the study of open quantum systems \footnote{A variant approach for open systems utilising a bi-Lanczos algorithm was recently developed in \cite{Bhattacharya:2023zqt}.}. Note that in the time-dependent case, as the evolution operator is not simply $\exp (-i t H )$ but is given instead by Dyson's time-ordered exponential, the Liouvillian is not determined simply in terms of the Hamiltonian. For periodically-driven dynamics, there is a way to bypass this issue for doing the Krylov construction.

\subsection{States}

Suppose that the initial state of a Floquet system (with time periodicity $T$) is $\ket{\psi(0)}$. We initiate the Krylov construction in the following manner. Let $U_F$ be the Floquet operator of the system and construct the Krylov subspace:
\begin{align}
\mathcal{H}_K &=\{\ket{\psi_0}, U_F \ket{\psi_0}, U_F^ 2 \ket{\psi_0},......\} \nonumber
\\ &=\{\ket{\psi_0}, \ket{\psi_1}, \ket{\psi_2},......\} \, ,
\end{align}
where $\ket{\psi_j}$ denotes the state of the system after a time $t=jT$. As is usual in the study of Floquet dynamics, we will only probe the system stroboscopically at these discrete times \footnote{For {\it kicked} Floquet sytems which we focus on in the next section,  the system evolves freely with no potential in between the localised kicks. The K-complexity growth in between two kicks will thus be that of a free integrable system.}. The Krylov subspace is thus naturally generated with the Liouvillian $\mathcal{L}=U_F$ by picking the maximal linearly independent set of vectors from the above sequence.
Note that since $U_F$ is unitary and not Hermitian, the usual Lanczos algorithm discussed in the previous section for constructing the orthonormal basis in this subspace will not work \footnote{One can check explicitly that the basis vectors generated in this way will not be orthogonal.}. 

We can however use a generalised `Lanczos' algorithm - known as the Arnoldi iteration - for constructing an orthonormal basis for the Krylov subspace. For this we proceed as follows. Define $\ket{K_0}=\ket{\psi_0}$ and
\begin{align} 
\ket{K_1}=\frac{1}{h_{1,0}}\big( U_F \ket{K_0}-h_{0,0}\ket{K_0} \big).
\end{align} 
Subsequently we continue recursively,
\begin{align}
\ket{K_n}=\frac{1}{h_{n,n-1}}\big( U_F \ket{K_{n-1}}-\sum_{j=0}^ {n-1}h_{j,n-1}\ket{K_j} \big).
\end{align} 

Here $h_{j,k}=\bra{K_j}{U_F}\ket{K_k}$. The $h_{n,n-1}$ are the Arnoldi coefficients which are given by the norm of the vector in the r.h.s numerator above. As before, the algorithm terminates if a null vector is generated \footnote{Note that for Lanczos iteration, $\ket{K_n}$ involves explicitly only $\ket{K_{n-1}} and \ket{K_{n-2}}$, while Arnoldi iteration requires all previous vectors as well.}. One can check that this Gram-Schmidt-like procedure, constructs an orthonormal basis in the Krylov subspace.

With the Krylov basis generated, the state complexity is defined as before: $\mathcal{K}_j^{\psi}=\bra{\psi_j}{\hat{K}}\ket{\psi_j}$ where the complexity operator $\hat{K}$ is defined in eq. \eqref{KO}. Since $\ket{\psi_j}=U_F^j\ket{K_0}$, this can be written as
\begin{align}   
\mathcal{K}_j^{\psi}=\sum_{n=0}^{D_K-1} n \, |\bra{K_n}{U_F^ j}\ket{K_0}|^2.
\end{align}
Similarly, the K-entropy after $j$ kicks is given by
\begin{align}
S_j^{\psi}=- \sum_{n=0}^{D_K-1} |\bra{K_n}{U_F^ j}\ket{K_0}|^2 \log |\bra{K_n}{U_F^ j}\ket{K_0}|^2.
\end{align}
These explicit expressions can be used to compute the state K-complexity and K-entropy once the Krylov basis and Floquet matrix are known. We will use them in the next section to investigate the time-dependence of these quantities in a specific model - the toral QKR.

\subsection{Operators}
  Operator complexity for studying operator growth is defined in a similar manner. Let $O_0$ be the operator under study at $t=0$ and construct
  \begin{align}   
  \mathcal{H}_K^O&= \{|O_0), |U_F^{\dagger}O_0 U_F),   |(U_F^{\dagger})^ 2 O_0 U_F^ 2),....\} \nonumber\\
  &=\{|O_0), |O_1), |O_2), .... \}\, ,
  \end{align} 
where $O_j$ denotes the Heisenberg picture operator at time $t=jT$. As before, $|O)$ denotes the operator $O$ as a ket in the linear space of all operators that act in $\mathcal{H}$. Note that here $U_F$, as a superoperator, has the action $U_F|O)=|U_F^ {\dagger} O U_F)$.

The Krylov basis is then generated by the following recursive algorithm. Define $|K_0)=|O_0)$ and
\begin{align}  
&|K_1)=\frac{1}{h_{1,0}}\big[ U_F|K_0)-h_{0,0}|K_0) \big] \\ 
&|K_n)=\frac{1}{h_{n,n-1}}\left[ U_F |K_{n-1})-\sum_{j=0}^ {n-1}h_{j,n-1}|K_j) \right] \label{Kn}
\end{align}  
with
  $h_{j,k}=(K_j|U_F|K_k)=\dfrac{1}{D}Tr(K_j^{\dagger} U_F^{\dagger}K_{k}U_F )$. As before, the normalisation $h_{n,n-1}$ are the Arnoldi coefficients.

With $O_j=(U_F^{\dagger})^j O_0U_F^ j$ being the time evolved operator at (stroboscopic) time $t=jT$, the operator complexity, defined by $\mathcal{K}_j^{O}=(O_j|\hat{K}|O_j)$, is given by
\begin{align}\label{KCFdefn}
\mathcal{K}_j^{O}&=\sum_{n=0}^{D_K-1} n |(K_n|U_F^ j|K_0)|^2 \nonumber\\
&= \frac{1}{D^ 2}\sum_{n=0}^{D_K-1} n |Tr[K_n^{\dagger} U_F^{\dagger j} K_0 U_F^j]|^2.
\end{align}
As before, this provides us a direct way to compute operator complexity given the Floquet matrix and the Krylov operator basis.
Using $|O_{j})=\sum_n \phi_n^j |K_n)$, where $\phi_n^j=(K_n|O_j)$ is the $n$-th operator amplitude at time $t=j T$, we can write an equivalent form of the above equation
\begin{equation}\label{KCFdefn2}
\mathcal{K}_j^{O}=\sum_n  n \, |\phi_n^j|^2\,.
\end{equation}

\subsection{Floquet dynamics on the Krylov chain}

Similar to the Krylov chain picture in terms of a particle-hopping model generated by the Lanczos procedure (briefly reviewed above in section \ref{overview}), there is an  analogous picture when the Arnoldi procedure is used to study the stroboscopic time-evolution for Floquet dynamics. We start with
\begin{align}
|O_{j+1})=U_F |O_j).
 \end{align}
Next we use $|O_{j})=\sum_n \phi_n^j |K_n)$ and eq. \eqref{Kn} which determines the action of $U_F$ on $|K_n)$. This leads to the following set of equations for the operator amplitudes
\begin{align}
\phi_0^{j+1}&= h_{0,0}\phi_0^j+h_{0,1}\phi_1^j+h_{0,2}\phi_2^j+....+h_{0,D_K-1}\phi_{D_K-1}^j \nonumber\\
\phi_1^{j+1}&= h_{1,0}\phi_0^j+h_{1,1}\phi_1^j+h_{1,2}\phi_2^j+....+h_{1,D_K-1}\phi_{D_K-1}^j \nonumber\\
\phi_2^{j+1}&= h_{2,1}\phi_1^j+h_{2,2}\phi_2^j+h_{2,3}\phi_3^j+....+h_{2,D_K-1}\phi_{D_K-1}^j \nonumber\\
\vdots \nonumber\\
\phi_{D_K-1}^{j+1}&= h_{D_K-1,D_K-2} \phi_{D_K-2}^ j+h_{D_K-1,D_K-1} \phi_{D_K-1}^ j 
\end{align}
which can be written in the compact form\footnote{This equation holds also for states instead of operators and also if the dimension of the Krylov subspace $D_K$ is infinite.} 
\begin{align}
\phi_i^{j+1}&=\sum_{l=i-1}^{D_K -1} h_{i,l} \, \phi _l^j
\end{align}
Here $j=0,1,2,....$ denotes the number of kicks (time parameter) and $i,l$ range over the $D_K$ Krylov chain sites: $i,l=0,1,2,..., D_K-1$. Also, $h_{0,-1}=0$ and $\phi_n^ 0= \delta_{n0}$. Note that instead of a continuous time differential equation (eq. \eqref{KCh}) for the operator amplitudes, in this case we find a discrete time (stroboscopic) difference equation since we are dealing with Floquet dynamics. This set of difference equations provide an alternative (and computationally simpler) recursive route to evaluating the operator amplitudes (and thereby K-complexity using eq. \eqref{KCFdefn2}) given the initial data (at $j=0$). We have checked that our results for K-complexity in the next section   obtained using this technique match with those using the method described earlier in this section. 

Note that the above equation also makes it clear that we have gone beyond nearest neighbour hopping, as distant Krylov chain sites are now involved. That longer range hopping is required when Arnoldi iteration is used to study Floquet dynamics in the Krylov chain picture was noted 
by \cite{Yates:2021asz}.

\section{K-complexity for quantum maps: the toral QKR} \label{TQKR}

As an illustration of our general method of Krylov construction for Floquet systems, we now investigate the dynamics of Arnoldi coefficients and K-complexity in a particular Floquet system, the QKR on a torus. Operator growth in quantum maps (although not Krylov complexity) has been studied previously in \cite{Sondhi:2018mvf}. For related studies of the OTOC, see \cite{Garcia-Mata:2018slr, Lakshminarayan:2018bge}.
The Hamiltonian of a particle (with mass $M$) on a circular ring with periodic delta function kicks (time period $T$) is
\begin{align} \label{HK}
H= H_0+V= \frac{p^ 2}{2 M}+ V(\theta)\delta_T(t)
\end{align}
 Here $\delta_T(t)= \sum_j \delta(\frac{t}{T}-j)$ and $\theta= 2 \pi q $, $q \in [0,1]$. For $V(\theta)= - \kappa \cos \theta / (2\pi)^2$, this defines the (classical and quantum) standard map on the cylinder, or the quantum kicked rotor which is a much studied system that is chaotic (for large enough $\kappa$).

The quantum system can be characterised also by the Floquet operator - the one-step time evolution operator relevant when the system is probed stroboscopically,
\begin{align}  
U_F=U_V U_0 \equiv \exp\big(-i T V(\theta)/\hbar\big)\exp\big(-i T P^ 2/2\hbar M\big).
\end{align}  

We will work with the simpler QKR on a torus, by imposing periodic boundary conditions on both $q$ and $p$. In position basis, the Floquet matrix for this quantum map (see \cite{lakshminarayan_2022} for a pedagogical discussion) takes the explicit form
   \begin{align*}
    (U_F)_{nn'} = \frac{1}{N}&e^{\frac{iN\kappa}{2\pi}\cos(2\pi\frac{(n+\alpha)}{N})}\\&\times\sum_{m=0}^{N-1}e^{\frac{-\pi i (m+\beta)^2}{N}} e^{\frac{2\pi i}{N}(m+\beta)(n-n')}
\end{align*}
Kinematically, this is Schwinger's finite dimensional model \cite{10.2307/70873} with both $Q$ and $P$ discrete with eigenvalues $q_n=(n + \alpha)/N$ and $p_m=(m + \beta)/N$, and has a finite dimensional Hilbert space. Here $N$ is the dimension of the Hilbert space, $\kappa$ is the strength parameter of the kicking and the matrix indices $n, n'$ run from $0$ to $N-1$. The $\alpha$ and $\beta$ parameters control the breaking of parity and time-reversal symmetries, respectively \footnote{Since we are primarily interested in how the dynamics is affected by changing the parameters $\kappa, \alpha, \beta$ we have set other parameters in the Floquet matrix such as $\hbar$, $T$ (time period of forcing), $R$ (radius of circle) and $M$ (mass of particle) to unity.}.

With the above explicit form of the Floquet Matrix and a given choice of the state/operator at $t=0$, the Krylov basis can be recursively generated and the K-complexity can be numerically computed as discussed in the previous section. The relevant results are discussed below\footnote{We obtained results for many set of values of $\kappa$, $\alpha$, $\beta$, $N$ and choice of the initial state/operator to ensure genericity. We are providing only some sample results here which illustrate the main qualitative features.}.

\subsection{Results}

\textbf{State Complexity:} To illustrate the principal features we choose a system size $N=350$ for the toral QKR and two sets of values of the coupling constants. See figs. \ref{fig:1}, \ref{fig:2}, \ref{fig:3} below.

\par \medskip

\textbf{Operator Complexity: } To study operator growth and complexity in the toral QKR, we again use two sets of values for the coupling constants and a system size of $N=32$ which gives the maximal Krylov space dimension $D_K^ {max}=993$. See figs. \ref{fig:4}, \ref{fig:5}, \ref{fig:6} below.
\par \medskip

We now discuss our principal results and observations and compare with similar results in the literature for time-independent systems.

In \cite{Dymarsky:2019elm} a relation between non-integrability and delocalisation on the Krylov chain was explored. In \cite{Rabinovici_2022, integrability_to_chaos} it was observed that (autonomous) integrable systems with enhanced symmetries/degeneracies show a stronger disorder in their Lanczos sequences as compared to chaotic ones and it was argued that this leads to a suppression of K-complexity for the integrable case. For similar observations regarding another chaos diagnostic, the out-of-time-ordered correlator, see \cite{Fortes:2019frf} and \cite{Garcia-Mata:2022voo} - the fluctuations in the OTOC after saturation are enhanced in the regular (weakly-coupled) limit as contrasted with the chaotic case. 

We clearly see a similar behaviour here for the Arnoldi coefficients (see figures \ref{fig:1} and \ref{fig:4}) whose fluctuations are much larger in the weakly coupled limit. Note that although we have presented the results for the Arnoldi coefficients $h_{n,n-1}$, which being normalisation coefficients are the analogues of the Lanczos coefficients, the same observations hold for the other Arnoldi coefficients \footnote{We have checked this for several other Arnoldi coefficients away from the sub-diagonal. We thank Pratik Nandy for asking this question for the diagonal $h_{n,n}$'s.}. Figures \ref{fig:2} and \ref{fig:5} (left) reveal how the size of the fluctuations in the Arnoldi coefficients, as measured by the standard deviation of the sequence $h_{n,n-1}$, decreases on increasing the system size. We note here that the plot for weak coupling is above the one for the chaotic case, signifying a larger magnitude of fluctuations in the weakly coupled case.

Also clear from figures \ref{fig:2} and \ref{fig:5} (right) is the sharp increase and saturation at the maximum possible value of the Krylov space dimension when the coupling constant is increased. Recall that for finite entropy quantum chaotic systems (with Hilbert space dimension $D$), \textit{generic} initial operators/states are expected to lead to a maximal Krylov space dimension ($D_K=D$ for states and $D_K=D^ 2-D+1$ for operators). The eigenvalues of the Liouvillian super-operator $\mathcal{L}$ are $\exp (i\epsilon_{ab})$ with $\epsilon_{ab} \equiv \epsilon_a - \epsilon_b$ ($\epsilon_a$ being a quasi-energy level). The zero phase (when $a=b$)
is always present and at least $D$ times degenerate \footnote{See section 2 of \cite{Rabinovici_2021} for a derivation of the bound on $D_K$ and where a discussion is provided on how
degeneracies in the spectrum lead to a lowering of the Krylov space dimension. The discussion there is for time-independent Hamiltonian systems but a similar argument holds for Floquet systems.} but additional degeneracies can arise due to the existence of symmetries or level crossings in the spectrum. In the case of figs. \ref{fig:2} and \ref{fig:5} (right) with fixed $\alpha, \beta$ and tunable coupling $\kappa$, the growth of the Krylov subspace dimension $D_K$ with $\kappa$ is due to the existence of level repulsion which is expected to arise for chaotic dynamics (for small values of $\kappa$  level crossings can give additional zero phases thus lowering $D_K$).

\begin{figure*}[htbp]
    \centering
    \includegraphics[height = 6.0 cm, width   =7.7cm]{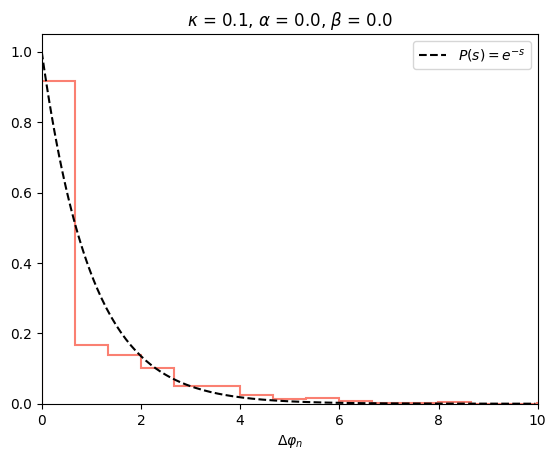}
    \includegraphics[height = 6.0 cm, width = 7.7 cm]{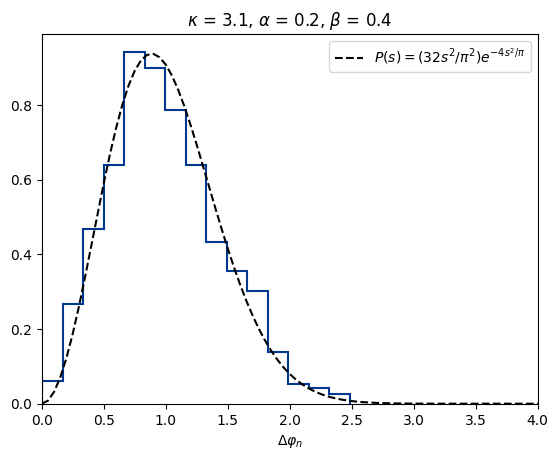}
    \caption{Spectral statistics of the quasi-energy level spacings in the integrable (\textbf{left}) and chaotic (\textbf{right}) cases.}
    \label{fig:8}
\end{figure*}

The K-complexity and K-entropy plots (\ref{fig:3} and \ref{fig:6}) also show a larger saturation value for the strongly coupled chaotic case. A large value of the K-complexity means a more efficient exploration of the Krylov chain (in the particle hopping model) which is expected for chaotic dynamics. Note that in our model no exponential growth in K-complexity is clearly discernible - this is expected as the scrambling time here is of the order of just a few kicks. However, the saturation of K-complexity does happen around the Heisenberg time ($t_H \sim D_K $).  Also plotted in fig. \ref{fig:8} are the spectral distributions for the quasi-energy level spacings in the integrable and chaotic cases. The two sets of parameter values chosen here are the same as for the operator K-complexity plots to facilitate comparison. As expected, in one case we get the Poisson distribution, whereas the other is Wigner-Dyson of the GUE universality class (both parity and time-reversal symmetries are broken in this case) and shows level repulsion  characteristic of chaos.

For these computations, to deal with issues of precision and pile-up errors, we used the full orthogonalisation numerical algorithm (see \cite{parlett1998symmetric} and  appendix C of \cite{Rabinovici_2021} for further details). 

As noted above, these observations hold for different system sizes, initial states and coupling constants. In fact, they also hold for other systems. An alternative finite-dimensional quantum map in which similar computations can be done is the kicked Harper map with the Floquet operator
\begin{align}
U_F= \exp\big(-i \frac{\kappa}{2\pi} \cos (2\pi Q)\big)\exp\big(-i \cos (2 \pi P)\big)
\end{align}
We have checked that the various plots obtained for this model, and related observations, are similar to the ones for the toral QKR.

\begin{figure*}
    \centering
    \includegraphics[height = 5.5 cm, width = 7.7cm]{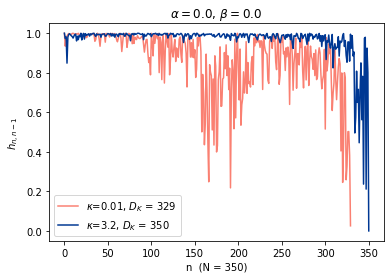}
    \includegraphics[height = 5.5 cm, width = 7.7cm]{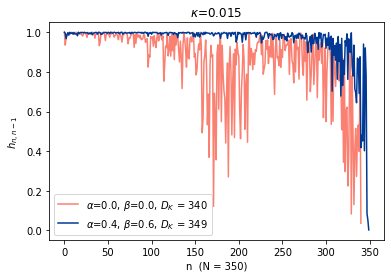}
    \caption{Dynamics of Arnoldi Coefficients. The plots highlight the larger fluctuations and smaller Krylov space dimension for weak coupling. The plot on the left shows the variation of $h_{n,n-1}$ with $n$ for $\alpha=\beta=0$, and a small and a large value of the coupling $\kappa$. The right plot is for the same quantities but now  for a fixed value of $\kappa$ and two sets of values for $\alpha, \beta$.}
    \label{fig:1}
\end{figure*}

\begin{figure*}[htbp]
    \centering
    \includegraphics[height = 5.0 cm, width = 7.5cm]{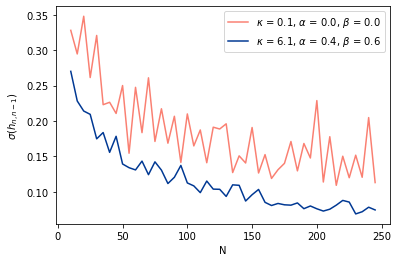}
    \includegraphics[height = 5.5 cm, width = 8.4cm]{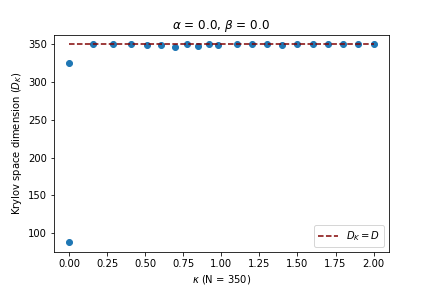}
    \caption{\textbf{(Left)} This plot shows how the fluctuation magnitude, as measured by the standard deviation, in the Arnoldi coefficients varies with system size ($N$). \textbf{(Right)} Krylov Space Dimension ($D_K$) vs. coupling strength. Note that $D_K$ increases with the coupling $\kappa$ and saturates quickly at the maximum possible value (dashed horizontal line).}
    \label{fig:2}
\end{figure*}

\begin{figure*}[htbp]
    \centering
    \includegraphics[height = 5.5cm, width = 7.7cm]{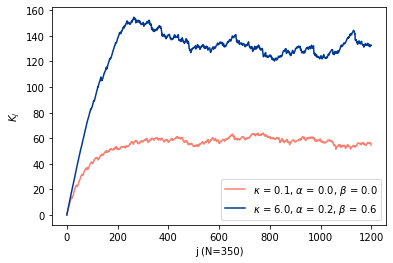}
    \includegraphics[height = 5.5 cm, width = 7.7cm]{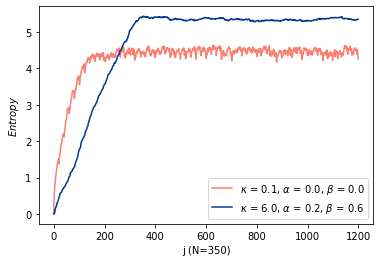}
    \caption{\textbf{Left}: State Complexity dynamics. Note the differing saturation behaviour for small and large coupling parameters.  
    \textbf{Right}: K-entropy growth with time (number of kicks).}
    \label{fig:3}
\end{figure*}

\begin{figure*}[htbp]
    \centering
    \includegraphics[height = 5.5 cm, width = 7.7cm]{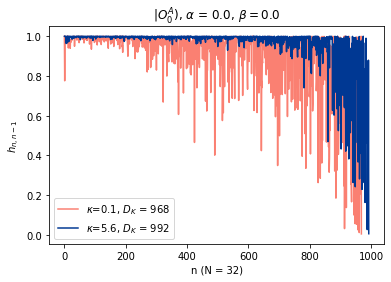}
    \includegraphics[height = 5.5 cm, width = 7.7cm]{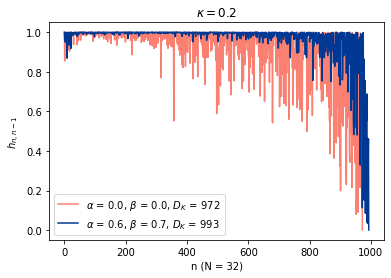}
    \caption{Dynamics of Arnoldi Coefficients. The plots highlight the greater magnitude of fluctuations in the weak-coupling limit.  Also to be noted is the smaller Krylov space dimension in this case.}
    \label{fig:4}
\end{figure*}

\begin{figure*}[htbp]
    \centering
    \includegraphics[height = 5.0 cm, width = 7.7cm]{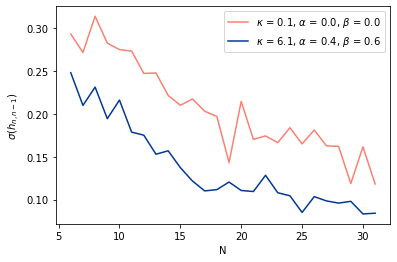}
    \includegraphics[height = 5.5 cm, width = 7.7cm]{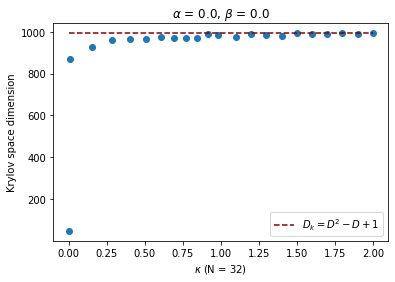}
\caption{\textbf{(Left)} Fluctuations versus system size. The plot shows how the size of fluctuations changes as the system size ($N$) is increased. There is the expected overall decrease, although it is not monotonic. \textbf{(Right)} Krylov Space Dimension $D_K$ vs. coupling $\kappa$. This plot makes manifest the chaotic nature of the dynamics as the coupling $\kappa$ is increased. $D_K$ increases rapidly to its maximum possible value.}
    \label{fig:5}
\end{figure*}

\begin{figure*}[htbp]
    \centering
    \includegraphics[height = 5.0 cm, width   =7.7cm]{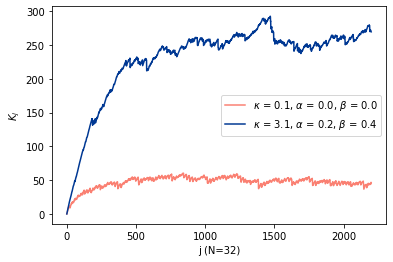}
    \includegraphics[height = 5.0 cm, width = 7.7 cm]{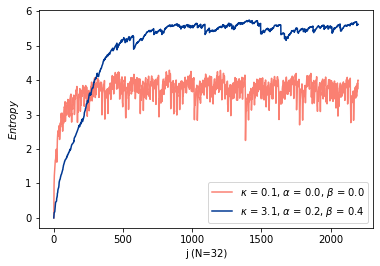}
    \caption{ \textbf{Left}: Operator Complexity dynamics. This plot shows the variation of K-complexity with time (number of kicks) for the weak and strong coupling cases. \textbf{Right}: K-entropy vs. the number of kicks.}
    \label{fig:6}
\end{figure*}

\section{Discussion} \label{summary}

In this paper we studied the Krylov construction and complexity for time-dependent systems. We give below a point-wise summary of the principal results of this paper.

\begin{itemize}
    \item 

We develop a general method for doing the Krylov construction for periodic time-dependent systems. The construction utilises the Floquet matrix and the Arnoldi iterative procedure for constructing the Krylov basis. 

\item
In our study of the toral QKR, we noticed that the fluctuations in the Arnoldi coefficients are much larger in the weak-coupling regime as compared to the chaotic case. This is not unexpected and, as noted above, is the case for other chaos diagnostics and for other systems as well. The larger fluctuations lead to localisation on the Krylov chain \cite{Rabinovici_2022, Dymarsky:2019elm}. We also studied the change in the magnitude of the fluctuations (as measured by the standard deviation of the sequence $\{ h_{n,n-1} \}$ and other Arnoldi coefficients) as the system size increases.  

\item
Likewise, we noted the distinctive behaviour of K-complexity of the toral QKR for weak/strong coupling - in terms of the late-time saturation in the two cases. A suppressed saturation value for the integrable case is in accord with results in the literature for time-independent systems \cite{Rabinovici_2021, Rabinovici_2022, integrability_to_chaos}, where this behaviour is associated with an enhanced localisation on the Krylov chain due to a stronger disorder in the hopping amplitudes. We also plotted the spectral statistics in the two cases and found the expected behaviour (Poisson vs. Wigner-Dyson).


\item 
We also determined how the Krylov subspace dimension ($D_K$) increases as a function of the coupling constant of the toral QKR. To the best of our knowledge, a detailed investigation of how $D_K$ changes on increasing the coupling strength in the Hamiltonian of a system has not been undertaken before in the literature.

\item Although for a concrete example we focussed on the toral QKR, we emphasise that our method is not limited to this system. For any Floquet system, not necessarily a delta function kicked one, we can use the same method. Of course, for other Floquet systems the Floquet matrix may not have a simple analytical form.

\item In the appendix, we carry out the analysis for the toral QKR using the standard Lanczos approach. Here we work with the effective Hermitian Hamiltonian $i \log U_F$. The observations regarding larger fluctuations in the Lanczos coefficients in the regular limit, and late-time saturation value of K-complexity, continue to hold in this approach as well.
\end{itemize}
There are some interesting directions to be pursued further.

\begin{itemize}
\item
It would be interesting to do this analysis for the canonical QKR (standard map with cylindrical phase space). In part this is because the classical-quantum correspondence is more readily studied in this model as there is a classical Hamiltonian which is canonically quantized (this is not the case for the finite $N$ toral QKR - an intrinsically quantum, finite dimensional system). The Floquet matrix in this case is, in the momentum eigen-basis, an {\it infinite} dimensional (banded) matrix, so this numerical study would be more challenging.

\item 
It remains to be seen if there is a useful way to construct the Krylov basis and define K-complexity for more general time-dependent systems. One interesting direction to pursue here would be to compute K-complexity for quenched dynamics. See \cite{Afrasiar:2022efk, Pal:2023yik, Gautam:2023pny} for some recent work in this direction.
\end{itemize}

We hope to report on these in the near future.


\section*{Acknowledgements}
AAN would also like to thank the Department of Physical Sciences, IISER Mohali, India and the organisers of the 5th Bangkok workshop on Discrete Geometry, Dynamics and Statistics at the Department of Physics, Chulalongkorn University, Bangkok, Thailand, where some of this work was first presented.  We would like to acknowledge the use of the computing facility at the physics department, Ashoka University. We would like to thank Somendra Bhattacharjee, Philip Cherian and Peng Zhao for useful discussions. AAN would like to especially thank Johannes Hofmann for carefully going through the manuscript and giving very detailed and helpful feedback and suggestions. We thank Anatoly Dymarsky and Pratik Nandy for interesting comments and questions, and pointing out some relevant references. We are also thankful to the referees for numerous suggestions and pointing out errors and inaccuracies.

\appendix

\section{Krylov construction with $H_F \equiv i\log U_F$} \label{lnU}

In this paper, since the Krylov construction involved a Liouvillian generated by the unitary Floquet operator $U_F$, we had to use the Arnoldi iteration as the Lanczos algorithm does not work in the non-Hermitian case. However, if we define $U_F \equiv \exp(-iTH_F)$, then the Floquet effecive  Hamiltonian $H_F \equiv i \log U_F$ (setting $T=1$) is Hermitian \footnote{Note that this is not the same as the actual Hamiltonian of the system given by eq. \eqref{HK}.} and can be used to initiate the usual Lanczos iteration based Krylov construction.  The time dependence of the Lanczos coefficients and operator complexity in this case (for the toral QKR with $N=32$) is given in the plots below. For Floquet spin chains, this method has been utilised in \cite{Yates:2021lrt, Yates:2021asz}.

\begin{figure}[H]
    \centering
    \includegraphics[height = 5.5 cm, width = 7.7cm]{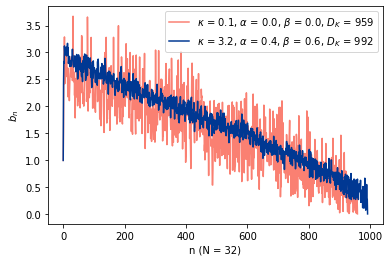}
    \includegraphics[height = 5.5 cm, width = 7.7cm]{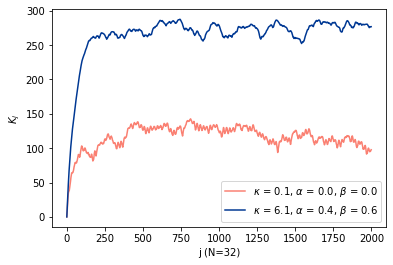}
    \caption{Lanczos Coefficients and Krylov Complexity generated by $\log U_F$.}
    \label{fig:7}
\end{figure}

We notice that the fluctuations in the Lanczos coefficients are enhanced for weak coupling, as before. Also, the K-complexity growth is similar to the Arnoldi analysis and the saturation values differ in the weak and strong coupling domain. 

This method, although it gives expected results for the coefficients and K-complexity, has some limitations \footnote{We thank the referee for a question regarding the merits of the two approaches.}.  In the Arnoldi approach with $U_F$, the basis generated by the Liouvillian is simple to interpret physically: the $j$th element is simply the state/operator after $j$ kicks. In the $\log U_F$ case, although one formally works with a Hermitian operator, this is not the system Hamiltonian - eq.\eqref{HK} - and its physical interpretation is unclear. There are also issues related to choosing the branch of the logarithm and folding of the spectrum which gives rise to ambiguities. There is a freedom in shifting quasi-energies by integer multiples of $2\pi$ which affects $b_n$’s.  For more detail, see the section on Krylov subspace dynamics in \cite{Yates:2021lrt}.

\DeclareFieldFormat*{title}{#1}
\DeclareFieldFormat{titlecase}{\textit{#1}}
\printbibliography


\end{document}